\documentclass{ws-jnmp}

\begin{document}

%\arttype{Letter} % default 'Article'

%\markboth{Authors' Names}
%{Instructions for Typing Manuscripts (Paper's Title)}

\markboth{N. Theodorakopoulos}
{Bubbles, clusters and denaturation in genomic DNA}
\catchline{X}{XX}{2011}{}{}
%
%%%%%%%%%%%%%%%%%%%%%%%%%%%%%%%%%%%%%%%%%%%%%%%%%%%%%%%%%%%%%%%%%%%%
\copyrightauthor{N. Theodorakopoulos}

\title{BUBBLES, CLUSTERS AND DENATURATION IN GENOMIC DNA: 
MODELING, PARAMETRIZATION, EFFICIENT COMPUTATION
%\\USING COMPUTER SOFTWARE\footnote{For the title, try not to use more than 3 lines. Typeset the title in 11~pt Times roman, uppercase and boldface.}
}

\author{\footnotesize NIKOS THEODORAKOPOULOS%\footnote{Typeset names in 9~pt roman, uppercase. Use the footnote toindicate the present or permanent address of the author.}
}

\address{Theoretical and Physical Chemistry Institute,  National Hellenic Research Foundation\\
Vasileos Constantinou 48, 116 35 Athens, Greece\\
and\\
Fachbereich Physik, Universit\"at Konstanz, 78457 Konstanz, Germany
%University Department, University Name, Address\\ City, State ZIP/Zone, Country
%\footnote{State completely without abbreviations, the affiliation and mailing address, including country. Typeset in 9~pt italic.}
\\
\email{ntheodor@eie.gr ; Nikos.Theodorakopoulos@uni-konstanz.de%\footnote{Typeset author e-mail address in single line}
}}
\date{today}
%\author{SECOND AUTHOR}

%\address{Group, Laboratory, Address\\ City, State ZIP/Zone, Country\\ \email{author\_id@domain\_name}}

\maketitle

\begin{history}
\today
%\received{(Day Month Year)}
%\revised{(Day Month Year)}
%\accepted{(Day Month Year)}
%\comby{(xxxxxxxxx)}
\end{history}

\begin{abstract}
The paper uses mesoscopic, non-linear lattice dynamics based (Peyrard-Bishop-Dauxois, PBD) modeling to describe thermal properties of DNA below and near the denaturation temperature. Computationally efficient notation is introduced for the relevant statistical mechanics. Computed melting profiles of long and short heterogeneous sequences are presented, using a recently introduced reparametrization of the PBD model, and critically discussed. The statistics of extended open bubbles and bound clusters is formulated and results are presented for selected examples.
\end{abstract}


%\ccode{2000 Mathematics Subject Classification: 22E46, 53C35, 57S20}

\section{Introduction}	
Thermal denaturation, i.e. separation of the two strands as a result of heating, is one of the oldest established physico-chemical results pertaining to the DNA molecule. The \lq\lq melting\rq\rq- of DNA was first observed \cite{Thomas54} very soon after  the determination of the double helical structure. Theoretical descriptions in the sixties, proposed by Poland and Scheraga (PS) \cite{PoSche, Fi}, largely based on the concept of the helix-coil transition \cite{ZimmBragg}, were subsequently refined and developed in considerable detail, incorporating known enthalpic data on controlled oligomer thermodynamics and, when combined with appropriate software, can claim significant predictive power in regard to actual experimental melting profiles \cite{WaBe85, Blake91}. Helix-coil models adopt a mesoscopic description of DNA, describing the open (coil) or closed (helix) state of an individual base pair in terms of a discrete, Ising-type  variable. By virtue of their construction they cannot describe any dynamic phenomena.

An interesting alternative was proposed two decades ago by Peyrard, Bishop and Dauxois (PBD) \cite{PB, dauxpeyr1}. In accordance with contemporary soft mode concepts related to structural phase transitions, they attempted a reduced, also mesoscopic, lattice-dynamics motivated description of the double helix. In their case, the relevant degree of freedom is a transverse displacement which represents how far an individual base pair is from equilibrium. This is locally determined by a nonlinear (typically Morse-like) potential which accounts for the effective hydrogen bonding linking bases together. The motion of neighboring base pairs is coupled in a way which favors double-helical ordering (DNA stacking interaction). As it turns out, this minimal nonlinear lattice-dynamics  (also proposed in the context of wetting phenomena \cite{KroLip}) generates one of the simplest known one-dimensional models with short-range interactions which exhibit an exact phase transition in the  thermodynamic limit.  

The PBD model of DNA does more than a \lq\lq demonstration of principle" regarding the denaturation transition. As shown by Cule and Hwa \cite{CuleHwa} the randomness introduced by base-pair heterogeneity results in a structured rounding of the transition (multistep melting) in qualitative accordance with experimental observation. Recently, using improved computational methods and a new, global set of model parameters, it has been possible to compute detailed, multipeak melting profiles of long DNA chains using only sequence information and salt concentration as input; although further parameter optimization is probably needed, agreement  with experiment is impressive \cite{NTh2010}.
The model has also been used to calculate neutron scattering structure factors, also in very good agreement with recent experimental results \cite{Wildes}.  Furthermore, as will be noted in this work, the low-frequency optical phonons of the model are in the range where Raman spectroscopy has located some vibrational activity in DNA \cite{Urabe81}.  

Important issues remain open. One of the key advantages of the PBD approach is its potential ability to describe local openings (\lq\lq denaturation bubbles") of the double helix, such as those known to occur during the initial stage of the transription process. The relatively slow dynamics involved in these events would be consistent with the coarse graining implicit in the model. Bubble statistics has been extensively studied in the framework of the PBD model, and some specific correlation with transcription initiation sites has in fact been claimed\cite{Choi04} and debated \cite{PeyrComm_etc,BishComm}. On a more modest - and yet quite  fundamental - level, the formation of a single bubble has been observed  by studying the melting of specially designed oligomers \cite{Montri}. As the results reported here will show, although the process and its rough temperature dependence are correctly accounted for by the PBD model, important details, notably the asymmetric biphasic character of the profile, are missing. This may be due to the fact that the global parameter set used in PBD modeling is either (i) not yet fully optimized or (ii) lacking in detail, e.g. in not taking proper account of the variations in stacking interaction between different sets of neighboring base pairs. 

The paper is structured as follows: Section 2 describes the model and introduces system parameters to be used. Section 3 presents thermodynamics, including some necessary background material, and focuses on (i) issues related to finite chains and (ii) heterogeneity. Section 4 presents computed melting profiles for both long and short chains and compares theoretical vs. experimental results. Section 5 discusses the statistical properties of extended objects, i.e. bubbles and clusters consisting, respectively, of consecutive open and bound base pairs. Section 6 summarizes results and perspectives.

\section{Model, Notation, Parameters}
I will use the version of the PBD model proposed in \cite{dauxpeyr1}. The total potential energy of a chain of $N$ base pairs can be represented as a sum of local and nonlocal contributions
\begin{equation}
	H_P = \sum_{j=1}^{N-1} W(y_{j},y_{j+1})  + 
 \sum_{j=1}^{N} V_j(y_j) \-,
\label{eq:Ham}
\end{equation}
where the transverse coordinate $y_j$ represents the separation of the two bases at the $j$th site, the local term, an on-site Morse potential
\[ V_j(y_j) = D_j(1-e^{-\alpha_j y_j})^2\]
describes the combined effects of hydrogen-bonding, stacking and solvent acting on the $j$th base pair, and the anharmonic elastic term 
\begin{equation}
	W(y_j,y_{j+1}) = \frac{k}{2}\left[ 1 + \rho e^{-b(y_j+y_{j+1})}  
\right](y_j-y_{j+1})^2
\label{eq:stack}
\end{equation}
models the nonlinear base-stacking interaction. The parameter $k$ describes the strength of the linear stacking interaction, i.e. the \lq\lq residual stacking\rq\rq\- which characterizes the disordered state. Most of the stacking energy which favors values $y_j \sim y_{j+1}\sim 0$ comes from the nonlinear term which is parametrized by its dimensionless strength $\rho$ and its range $1/b$. In view of the very large ratio of double-stranded and single-stranded DNA persistence lengths  \cite{elasticity}, 
a value of $\rho=50$ appears reasonable. A stacking interaction range $1/b=5 A$ is consistent with the double helical geometry (diameter $21 A$). The other parameters are the depth $D_j$ and the range $1/\alpha_j$ of the Morse potential. This work follows the choice of parameters made in \cite{NTh2010}, i.e. $k=0.45 \> {\rm meV}/A^2$, $\alpha_{GC}=6.9 A^{-1}$, $\alpha_{AT}=4.2 A^{-1} $ and 
\begin{eqnarray}
\label{eq:DATpred}
\nonumber
D_{GC}  &=&   0.1655 + 0.00615 \log \frac{c}{0.075} \quad \quad {\rm eV} \\
D_{AT}  &=&   0.1255 + 0.00855 \log \frac{c}{0.075}   \quad \quad {\rm eV} 
\end{eqnarray} 
where c is the molar salt concentration. It may be noted that although these particular parameters were derived by fitting melting profiles, the resulting optical frequencies (with an effective mass per site equal to 618 amu) are about $83$ cm$^{-1}$ for GT and $44$ cm$^{-1}$ for AT. These values compare favorably with the results of low-frequency Raman spectroscopy \cite{Urabe81} which identify a broad band below 85 cm$^{-1}$.

\section{Thermodynamics}
\subsection{Definitions}
The thermodynamics of the model (\ref{eq:Ham}) is determined by the interplay between the elastic interaction and the Morse potential. At low temperatures, the displacements $\{y_n\}$ oscillate around the minima of the Morse potential, $y_n=0$; phonon spectra have a gap. At high temperatures displacements are large, oscillations occur on the flat top of the Morse potential, phonons are essentially gapless. As a result, thermodynamic properties are those of the harmonic chain. The classical partition function develops an infrared divergence, due to contributions from long-wavelength phonons. Some consequences of this divergence are fairly trivial, e.g. the emergence of a divergent temperature-independent, entropic contribution. Others, in particular when it comes to oligomers, are critical. In order to keep proper track of this divergence, I will introduce the size of the allowed phase space $L$ explicitly.    

The classical partition function of an open-ended chain with $N$ base pairs is thus defined by
\begin{equation} 
Z_N (L) = \int_{-\infty}^{L} dy_1 \cdots    \int_{-\infty}^{L}  dy_{N}\>  e^{- \beta H_P}   \quad,
\label{eq:Z}
\end{equation}
where $\beta=1/(k_B T)$, $k_B$ is the Boltzmann constant and $T$ the system temperature. Note that, because of the repulsive core of the Morse potential, contributions to the partition function for negative displacements decay quite rapidly. Thus, even though numerical computations always involve a lower cutoff $y_{min}$ of the integrations, if the cutoff is appropriately chosen, the values of the integrals remain practically independent of  the particular value of that cutoff.

A quantity which is of direct experimental interest is the fraction of bound (or, respectively, open) pairs. The microscopic average which is necessary to compute this is the probability that the $n$th base pair is bound, i.e.
\begin{equation} 
p_n = \frac{1}{Z_N (L)} \int_{-\infty}^{L} dy_1 \cdots   
 \int_{-\infty}^{y_c} dy_n \cdots   
 \int_{-\infty}^{L}  dy_{N}\>  e^{- \beta H_P}   \quad,
\label{eq:nthbp}
\end{equation}
where $y_c$ is a - somewhat arbitrary - crossover distance which distinguishes open from bound base pairs.  The calculations presented in this paper use a value $y_c=2$ A.

The fraction of open pairs is then given by
\begin{equation} 
\theta = 1 -\frac{1}{N} \sum_{n=1}^N p_n    \quad.
\label{eq:meltfrac}
\end{equation}

\subsection{Preliminaries. The homogeneous case}
\subsubsection{The transfer integral equation}
\begin{figure}[th]
%\centerline{\psfig{file=jnmpf1.eps,width=4.8cm}}
\centerline{\psfig{file=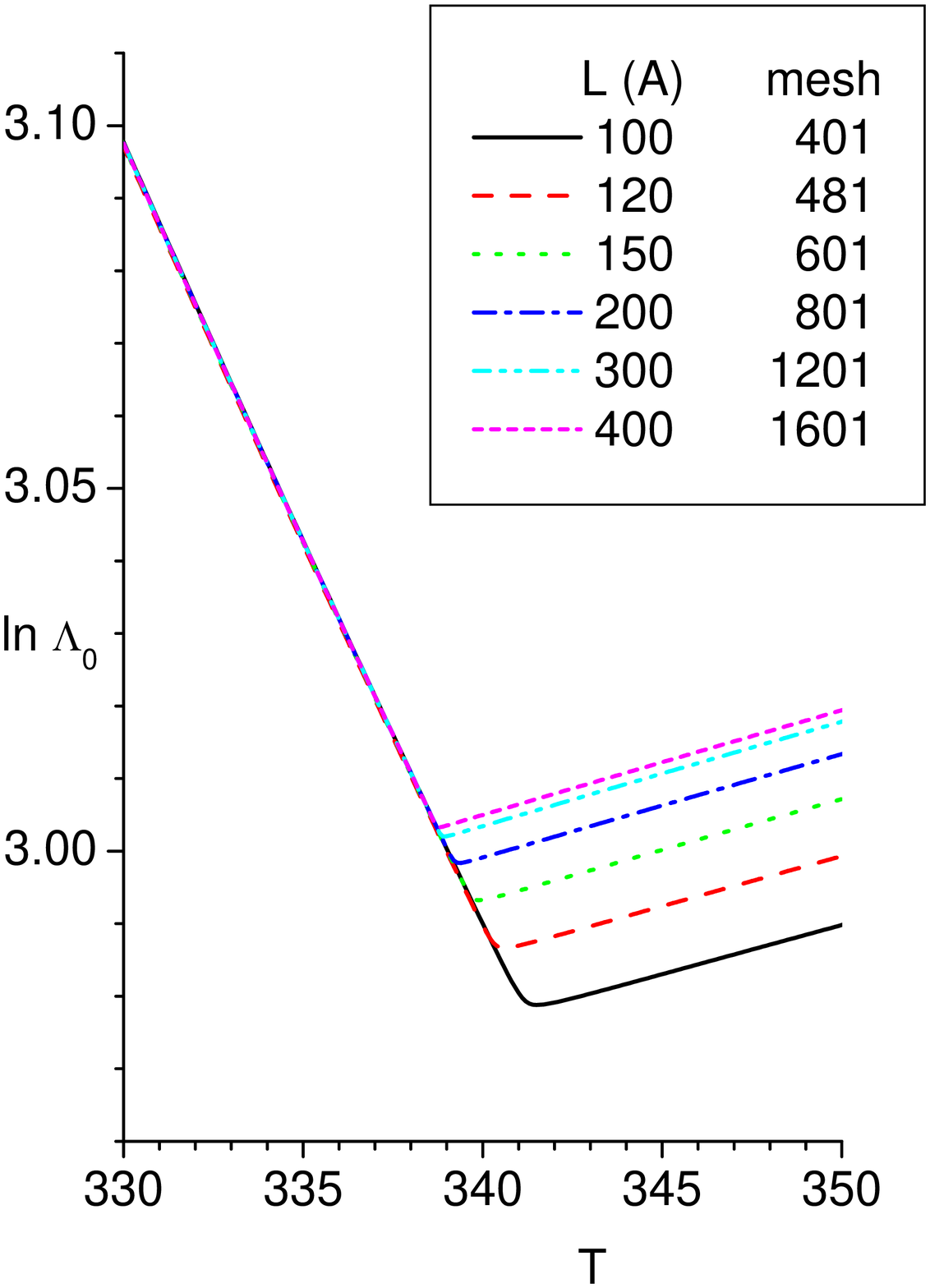,width=6cm}\psfig{file=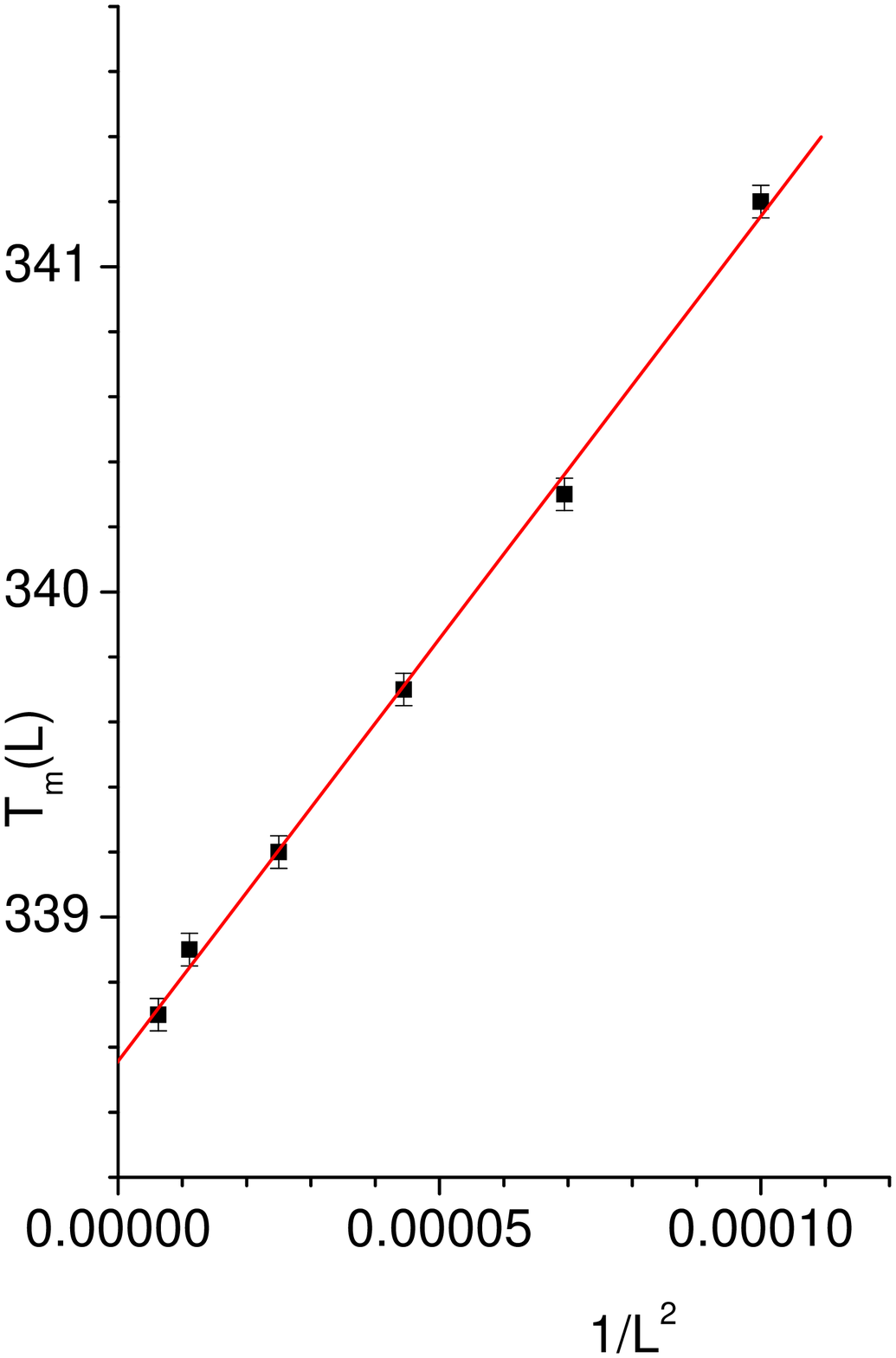,width=6cm}}
%\resizebox{\textwidth}{!}
%{\includegraphics{Lambda0.eps}\includegraphics{TmVsL.eps}}	
\vspace*{8pt}
\caption {The left panel shows the largest eigenvalue of (\ref{eq:TI}) as a function of temperature for a variety of upper cutoffs $L$. The right panel illustrates the dependence of the critical temperature (estimated by the position of the maximum of the second temperature derivative of $\Lambda_0$) on the cutoff $L$. 
}
\label{fig:eignum}
\end{figure}

In the homogeneous case 
\[D_j=D_{AT}, \alpha_j=\alpha_{AT},\> \forall j \] 
 the partition function can be calculated in terms of the spectra of the transfer integral equation,
\begin{equation}
	\int_{-\infty}^{L} dy'K(y,y') \phi_{\nu}^{(L)}(y') = \Lambda_\nu^{(L)} \phi_{\nu}^{(L)}(y)
\label{eq:TI}
\end{equation}
where
\begin{equation}
K(y,y')=     e^{-\beta W(y,y')} e^{-\beta [V_{AT}(y)+V_{AT}(y')]/2}	
\label{eq:TIkern}
\end{equation}
will be used in this work as a reference kernel, and the eigenfunctions are normalized to unity.
 Fig. \ref{fig:eignum} shows the numerically calculated lowest eigenvalues for different values of $L$. The discretization of the integral equation was performed using Gauss-Legendre quadratures in the interval $(y_{min}=-1.5 A,L)$, where $L$ varies between $100$ and $400 \> A$ and the density of the mesh has been kept constant at $4$ points per A.  The figure provides a numerical illustration of the existence of a well defined limit $L \to \infty$ of the integral equation (\ref{eq:TI}). Note in particular the rapid (quadratic) convergence of the successive numerical estimates of the critical temperatures to the limiting value.   
\subsubsection{The partition function}
The partition function (\ref{eq:Z}) can be written in terms of the spectra of (\ref{eq:TI}) as
\begin{equation}
	Z_N(L) = \sum_{\nu} \Lambda_\nu^{N-1}I_\nu^2 
\label{eq:ZTI}
\end{equation}
where the sum runs over all eigenstates, 
\[
I_\nu = \int_{-\infty}^{L} dy \> e^{-\beta V_{AT}(y)/2} \phi_\nu (y)
\]
and the $L$-superscripts have been dropped from eigenvalues and eigenfunctions for the sake of notational clarity. It should be understood that any eigenfunction and/or eigenstate is the result of a numerical calculation performed with a cutoff and on a grid.

The free energy per lattice site (base pair) is given by 
\begin{equation}
-\beta f_N(L,T)= \frac{1}{N}  \ln Z_N(L) =(1-\frac{1}{N}) \ln \Lambda_0
+ \frac{1}{N} \ln \left[   \sum_{\nu} 
\left(\frac{\Lambda_\nu}{\Lambda_0}\right)^{N-1}I_\nu^2 
\right]   \quad.
\label{eq:fTI}
\end{equation}
The above form is useful because it illustrates important limiting behavior. 
\subsubsection{The infinite chain}
In the thermodynamic limit, $N \to \infty$, terms of order $1/N$ in
(\ref{eq:fTI}) vanish. This kills all infrared-divergent terms and leaves only the contribution from the highest eigenvalue; for that contribution however, the limit $L \to \infty$ exists (Fig. \ref{fig:eignum}). The singularity of the spectrum of (\ref{eq:TI}) where the bound state merges into the continuum generates an infrared-free singularity of the thermodynamic properties of the infinite chain. The question of the order of the transition has been extensively discussed in the literature.
The dichotomy between apparent (first-order) and exact limiting (continuous) behavior has been presented in some detail in the case of the helicoidal version of the PBD model\cite{helicoidal}. The distinction is generic and applies to the model used here as well.
 
\subsubsection{The finite chain}
\begin{figure}[th]
%\centerline{\psfig{file=jnmpf1.eps,width=4.8cm}}
\centerline{\psfig{file=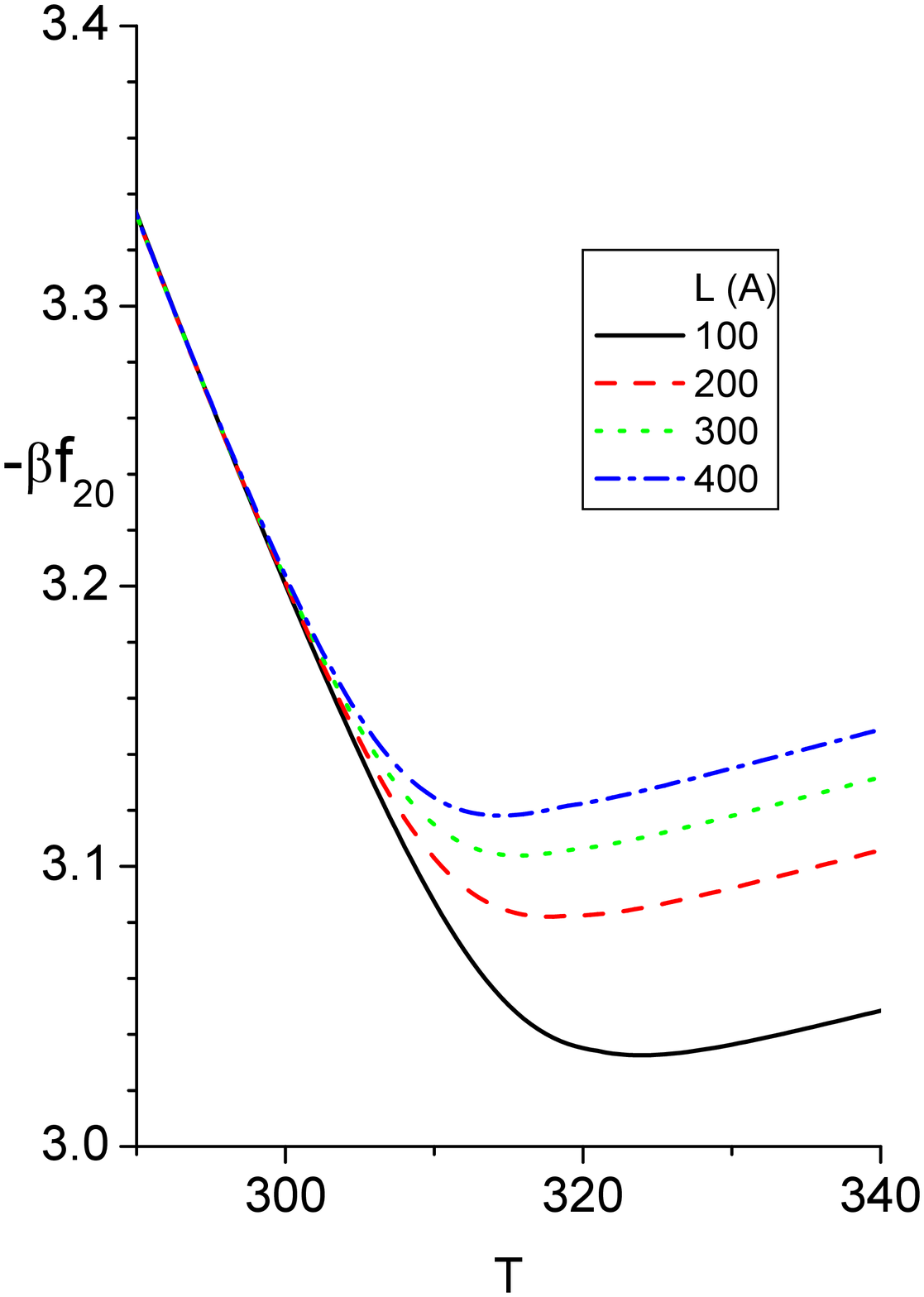,width=6cm},\psfig{file=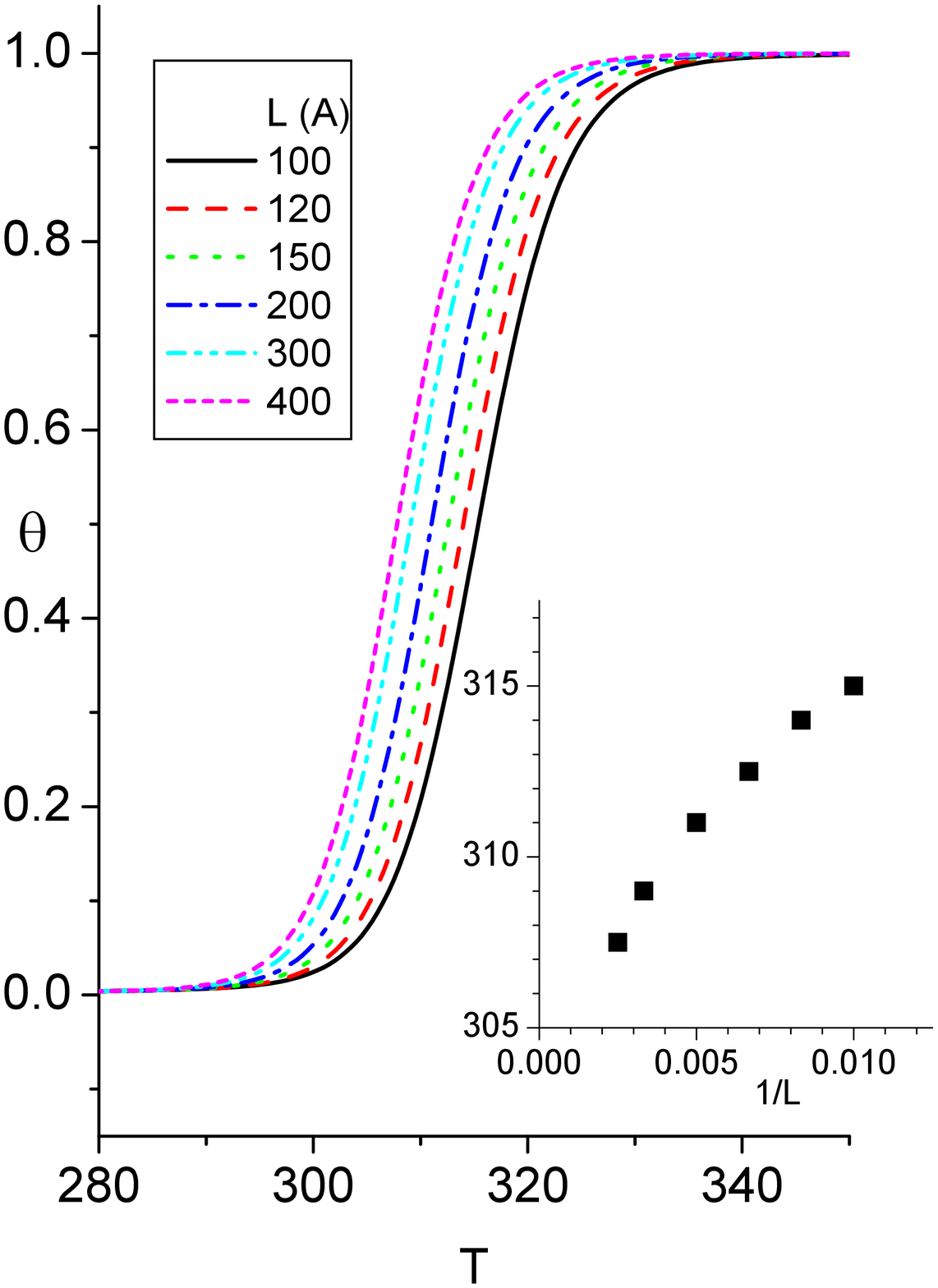,width=6cm}}
%\resizebox{\textwidth}{!}
%{\includegraphics{fren20Ldep.eps}\includegraphics{melt20Ldep.eps}}	
\vspace*{8pt}
\caption {The left panel shows the free energy per site (\ref{eq:fTI}) as a function of temperature for N=20 and a variety of upper cutoffs $L$. The right panel shows the melting fractions for N=20 and varying L; the inset illustrates the dependence of the critical temperature (estimated by the peak in the temperature derivative of the melting curve) on the cutoff $L$. 
}
\label{fig:n20}
\end{figure}

Fig. \ref{fig:n20}  (left panel) shows numerical results for the free energy per site in the case $N=20$. The transition is much more rounded, as would be expected from finite-size corrections. Moreover, the infrared divergent terms are significant on a visible scale. The concomitant effect on the melting profile can be seen on the right panel. There is no sign of convergence of the melting curve; as the inset indicates, the sequence of effective melting temperatures cannot be used in a straightforward way to produce a limit.
\subsubsection{The long chain}
The difficulties encountered in the previous section would encourage the following 
 \lq\lq fundamentalist\rq\rq argument regarding melting of any finite chains: since, for any finite $N$, the integral in (\ref{eq:nthbp}) remains finite in the limit $L \to \infty$, whereas the partition function diverges, $p_n$ should vanish at any nonzero temperature; it would follow that $\theta =1$ for any $T>0$.  

How watertight  is this formally correct \lq\lq proof"? 

In order to make this quantitative, let me go back to the free energy function. Suppose I would like the finite-size, infrared divergent corrections (cf. Fig. \ref{fig:n20}) to remain very small, e.g.
\begin{equation}
\frac{1}{N}\ln L < \epsilon  \quad,
\label{eq:criterion}
\end{equation}
where $\epsilon$ is some small number. 
On the other hand, one must note that the $L$'s used are not entirely arbitrary. $L$ must be large enough to ensure proper convergence of the TI equation (\ref{eq:TI}). For the set of parameters used here, it seems  that values between 300 and 400 A should be sufficient. One may then argue that if, in addition, (\ref{eq:criterion}) holds, the thermodynamic properties of the finite chain with values of $L$ in this order will be numerically stable, i.e. $L$-independent. What might happen at \lq\lq astronomic\rq\rq values of $L$ can have no practical bearing to DNA melting. 

\begin{figure}[th]
\centerline{\psfig{file=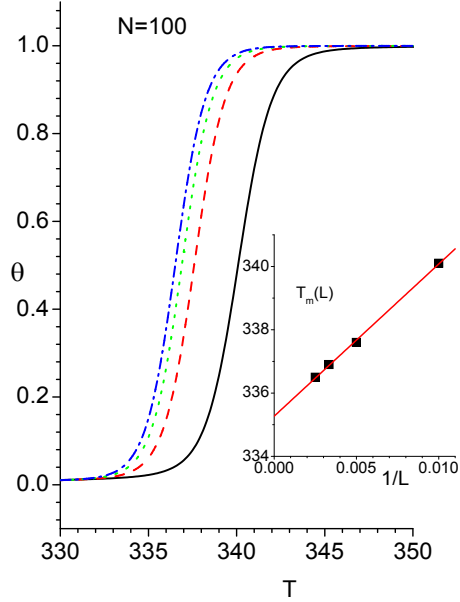,width=6.8cm}}
%\resizebox{0.5\textwidth}{!}
%\includegraphics{MeltProfN100AT.eps}	
\vspace*{8pt}
\caption {The melting fractions for N=100 as a function of temperature  and varying L; the inset illustrates the dependence of the critical temperature (estimated by the peak in the temperature derivative of the melting curve) on the cutoff $L$. 
}
\label{fig:n100}
\end{figure}

A typical choice of $\epsilon=0.01$ and $L=400$ A suggests that $N>600$ is a long enough chain to produce such numerically stable melting profiles within the PBD model. Interestingly, this is in line with what model-independent estimates based on analyzing melting temperatures of natural DNA classify as the limit of a long chain. This may however be too conservative a choice when it comes to PBD model considered here. Fig. \ref{fig:n100} shows that for $N=100$ the melting profiles are numerically stable in the sense described above and produce a linearly convergent sequence of melting temperatures.
\subsubsection{The double-stranded ensemble}
\label{seq:dss}
It is possible to define an infrared divergence-free thermodynamic average quantity by considering the conditional probability of any base pair being open, subject to the condition that the strands are not totally separated \cite{PeyrLop2009}. The use of such a \lq\lq double stranded ensemble\rq\rq  is standard procedure in analyzing DNA melting profiles of oligomers in terms of statistical (Poland-Scheraga type) models, because there the theory is entirely formulated in terms of this conditional probability (internal melting fraction). The probability of fulfilling this condition (external melting fraction), which is necessary in order to analyze experimental data, must then be obtained or estimated by other means. Within the PBD formalism the internal melting fraction can be calculated by first considering the (infrared-divergent) statistical weight corresponding to \lq\lq total melting\rq\rq, i.e. \cite{PeyrLop2009}.
\begin{equation}
Z_N^{*}(L)= \int_{y_c}^{L} dy_1 \cdots    \int_{y_c}^{L}  dy_{N}\>  e^{- \beta H_P}   \quad.
\label{eq:Zstar}
\end{equation}
The quantity
\begin{equation}
p_{ext}(L)= 1-\theta_{ext}(L)=1-\frac{Z_N^{*}(L)}{Z_N(L)}
\label{eq:pext}
\end{equation}
expresses the probability that at least one base-pair is bound. The conditional probability $p_n^*$ that the $n$th base pair is bound, provided that the chain is not entirely open, substitutes $Z_N(L)-Z_N^*(L)$ for $Z_N(L)$ in the denominator of
(\ref{eq:nthbp}) and should be free of infrared divergences. The same should hold for the internal melting fraction
\begin{equation}
\theta_{int}= 1-\frac{1}{N}\sum_{n=1}^N p_n^{*}   \quad.
\label{eq:thetaint}
\end{equation}

\begin{figure}[th]
\centerline{\psfig{file=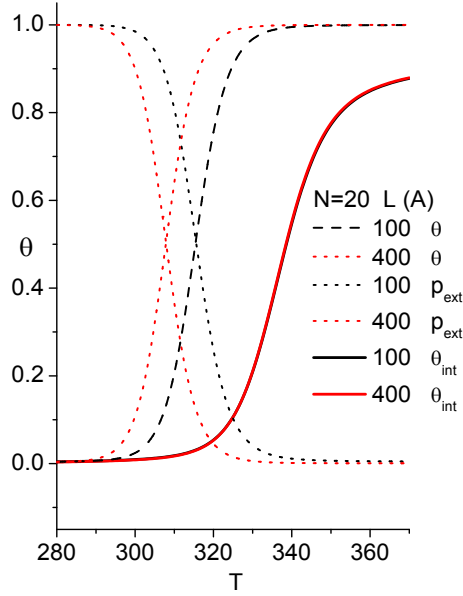,width=6.8cm}}
\vspace*{8pt}
\caption {The dashed curves represent melting fractions for a short (N=20), homogeneous AT chain, computed for $L=100, 400$. The corresponding external melting fractions are represented by the dotted curves. The solid curves (on top of each other) are the internal melting fractions. 
}
\label{fig:n20dse}
\end{figure}

Fig.  \ref{fig:n20dse} illustrates the use of the double-stranded ensemble for oligomers. The dashed curves are repetitions of the melting profiles of Fig    for $N=20,  L=100, 400$. The dotted curves represent $p_{ext}$ in the two cases. The solid curves, which coincide, are the internal melting fractions $\theta_{int}$.  

I will return to a constructive use of the double-stranded ensemble in Section \ref{seq:short}.

\subsection{Thermodynamics of the heterogeneous case}
\label{sec:heteroTherm}
\subsubsection{From multiple integrations to matrix products}
In the absence of translational invariance the partition function (\ref{eq:Z}) must be calculated directly as an $N$-dimensional integral; to do this in real-space, with a typical mesh of $1201$ points (which is what acceptable numerics demands, corresponding to an $L=300$ A) is an almost impossible task for $N>100$. This is not just a matter of computing time, but also of maintaining numerical accuracy. 

A useful alternative is to exploit the eigenfunction expansion of the reference (AT) kernel
\begin{equation}
K(y_j,y_{j+1})= \sum_{\nu_j} \Lambda_{\nu_j}
\phi_{\nu_j}(y_j)\phi_{\nu_{j}}(y_{j+1})
\label{eq:kernexp}
\end{equation}
in order to transform the  $N$-dimensional integration in real space to a matrix multiplication in a restricted eigenfunction space, e.g. by demanding that the $\nu$ summation be restricted to those states which satisfy $\Lambda_{\nu}/\Lambda_0<10^{-8}$. 
Using the auxiliary quantities
\begin{eqnarray}
\nonumber
A_{\nu}^{(j)}&=&\left(\frac{\Lambda_{\nu}}{\Lambda_0}\right)^{1/2} 
\int_{-\infty}^L dy \> \phi_{\nu}(y)e^{-\frac{\beta}{2} V_{AT}(y)-\beta \Delta V_j(y)   }
\quad j=1,N \\
%\nonumber
%C_{\nu_{N-1}}^{(N)}&=&\left(\frac{\Lambda_{\nu_{N-1}}}{\Lambda_0}\right)^{1/2} 
%\int_{-\infty}^L dy \> \phi_{\nu_{N-1}}(y)e^{-\frac{\beta}{2} V_{AT}(y)-\beta \Delta V_N(y)   }
%\\
B_{\nu_1 \nu_2}^{(j)}&=&\frac{ \left(\Lambda_{\nu_1} \Lambda_{\nu_2} \right)^{1/2}}{ \Lambda_{0}}
\int_{-\infty}^{L} dy \> \phi_{\nu_{1}}(y) \phi_{\nu_{2}}(y)
e^{-\beta \Delta V_j(y)} \quad j=2,\cdots, N-1,  
\label{eq:vectordefs}
\end{eqnarray}
where $\Delta V_j(y)  = V_{GC}(y)-V_{AT}(y)$ if $j$ is a GC site and 0 if $j$ is an AT site, it is straightforward to rewrite the partition function as 
\begin{equation}
Z_N(L) =  \Lambda_0^{N-1}   \sum_{\nu_1,\nu_2 \cdots \nu_{N-1} } 
A_{\nu_1}^{(1)}
B_{\nu_1 \nu_2}^{(2)}B_{\nu_2 \nu_3}^{(2)} \cdots
B_{\nu_{N-2} \nu_{N-1}}^{(N-1)}
A_{\nu_{N-1}}^{(N)}
\label{eq:Zmatr}
\end{equation}
or, in more compact matrix notation,
\begin{equation}
Z_N(L) =  \Lambda_0^{N-1}   
< {\bf A}^{(1)} |
{\bf B}^{(2)}  \cdots {\bf B}^{(N-1)}
| {\bf A}^{(N)}  > \quad.
\label{eq:Zmatr2}  
\end{equation}
Computing the probability of the $n$th base pair being in the bound state involves transforming the numerator of 
(\ref{eq:nthbp}) in a similar fashion.
I use script letters to define vector components similar to (\ref{eq:vectordefs}), with an upper limit equal to $y_c$, i.e.
 \begin{equation}
{\cal B}_{\nu_1 \nu_2}^{(j)} =\frac{ \left(\Lambda_{\nu_1} \Lambda_{\nu_2} \right)^{1/2}}{ \Lambda_{0}}
\int_{-\infty}^{y_c} dy \> \phi_{\nu_{1}}(y) \phi_{\nu_{2}}(y)
e^{-\beta \Delta V_j(y)}  
 \end{equation}
and similarly for ${\cal A}_{\nu}^{(1)}$. Then
\begin{eqnarray}
\nonumber
p_1& =&  
\frac  {
< {\bf {\cal A}}^{(1)} |
{\bf B}^{(2)}  
\cdots
 {\bf B}^{(N-1)}
| {\bf A}^{(N)}  >}
{< {\bf A}^{(1)} |
{\bf B}^{(2)}  \cdots {\bf B}^{(N-1)}
| {\bf A}^{(N)}  >
}\\
\nonumber
p_n &= & 
\frac  {
< {\bf A}^{(1)} |
{\bf B}^{(2)}  
\cdots
{\bf {\cal B}}^{(n)}
\cdots
 {\bf B}^{(N-1)}
| {\bf A}^{(N)}  >}
{< {\bf A}^{(1)} |
{\bf B}^{(2)}  \cdots {\bf B}^{(N-1)}
| {\bf A}^{(N)}  >
}
 \quad  n=2,\cdots,N-1 \\
p_N& =&  
\frac  {
< {\bf A}^{(1)} |
{\bf B}^{(2)}  
\cdots
 {\bf B}^{(N-1)}
| {\bf {\cal A}}^{(N)}  >}
{< {\bf A}^{(1)} |
{\bf B}^{(2)}  \cdots {\bf B}^{(N-1)}
| {\bf  A}^{(N)}  >
}   \quad.
\label{eq:pnmatr}  
\end{eqnarray}
\subsubsection{Further improvements in computational efficiency}
Computing $p_n$ for all sites involves heavy duplication of matrix multiplications. It is computationally much more efficient to define, compute and store intermediate results in vector form. Moreover, it is necessary to renormalize vectors to unity at every stage of a matrix-vector multiplication. The latter procedure minimizes numerical error because it keeps computed scalar products in the order of unity.

Let the unit vectors at the ends be defined via
\begin{eqnarray}
| {\bf  v}^{(1)}  > &=& \frac{1}{\mu_1^R} | {\bf  A}^{(N)}  >\\
< {\bf  u}^{(1)}  | &=&  < {\bf  A}^{(1)}  |\frac{1}{\mu_1^L}
\end{eqnarray}
where the $\mu$'s are the norms of the respective unnormalized vectors. I then store the result of each successive matrix-vector multiplication in vector form, as a unit vector and a norm, i.e.
\begin{eqnarray}
{\bf B}^{(N-j)}| {\bf  v}^{(j)}  > &=& \mu_{j+1}^R  | {\bf  v}^{(j+1)}  >\\ \nonumber
<{\bf u}^{(j)} | {\bf B}^{(j+1)}   &=&   <{\bf u}^{(j+1)}| \mu_{j+1}^L  \quad  j=1,2,\cdots, N-2.
\end{eqnarray}
It then follows that (i)
\begin{equation}
Z_N  = \Lambda_0^{N-1} \mu_1^L  \cdots  \mu_n^L  
\mu_{1}^R \cdots  \mu_{N-n}^R
< {\bf u}^{(n)}  | {\bf v}^{(N-n)}> 
\end{equation}
for any choice of $n=1,2,\cdots, N-1$, and (ii)
\begin{eqnarray}
p_n &=& \frac{1}{\mu_{n}^L} 
\frac { <{\bf u}^{(n-1)}|   {\bf {\cal B}}^{(n)}  |{\bf v}^{(N-n)}>  } 
{<{\bf u}^{(n)} |{\bf v}^{(N-n)}> }     \quad  n=2,\cdots, N-1
\nonumber \\
p_1   &=&   \frac{1} {\mu_{1}^L}  
\frac {<{\bf {\cal A}}^{(1)} |{\bf v}^{(N-1)}>}
{<{\bf u}^{(1)} |{\bf v}^{(N-1)}>} \nonumber \\
p_N   &=&   \frac{1} {\mu_{N}^R}  
\frac {<{\bf  u}^{(N-1)} |{\bf {\cal A}}^{(N)}>}
{<{\bf u}^{(N-1)} |{\bf v}^{(1)}>}    \quad.
\end{eqnarray}
Computing and storing the necessary intermediates for a melting profile of a chain with $N$ base pairs in vector form is a process with  ${\cal O}(N)$ computational steps.
\section{Computed melting profiles}
\subsection{Long chains}
\begin{figure}[th]
\centerline{\psfig{file=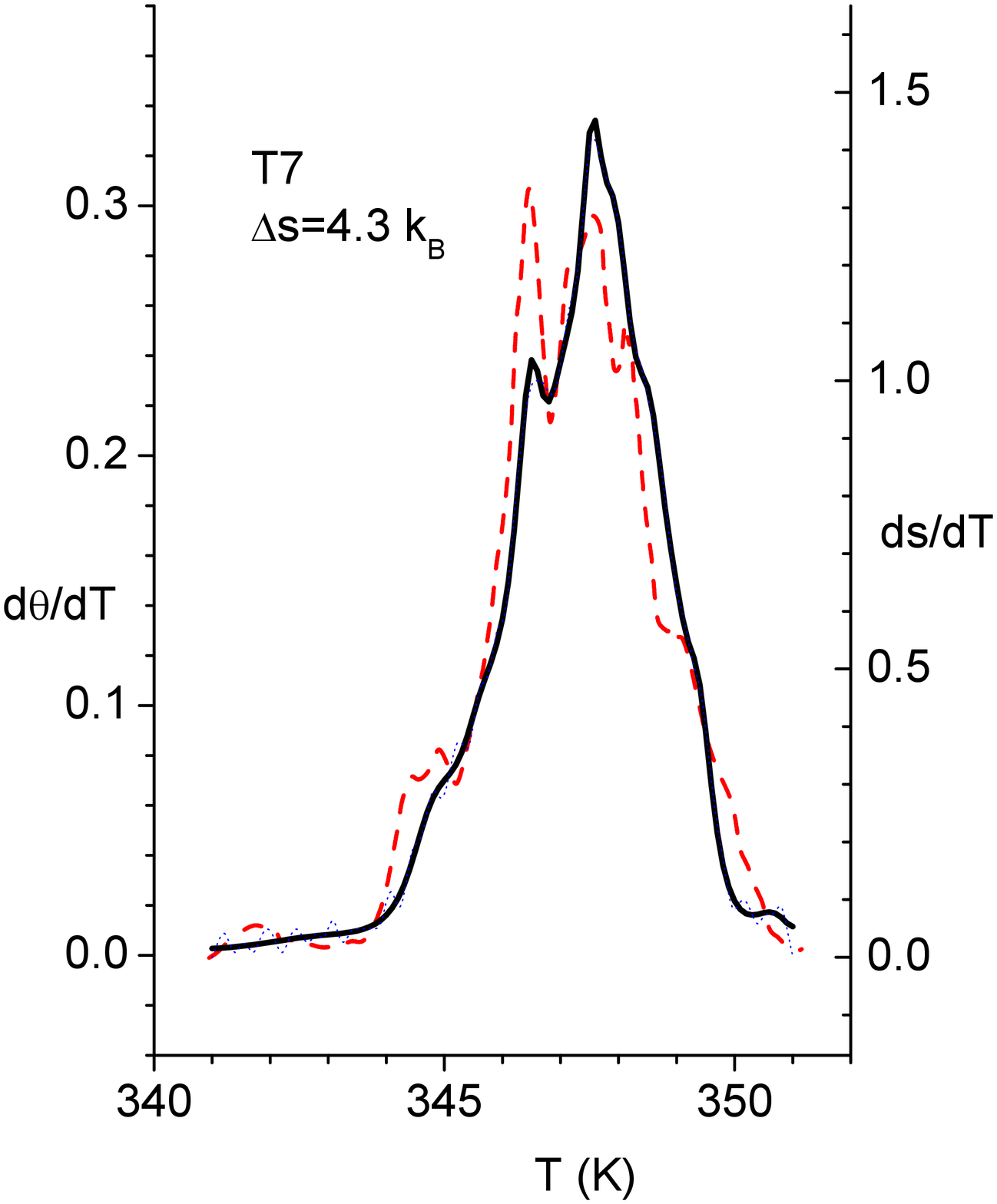,width=6cm} \psfig{file=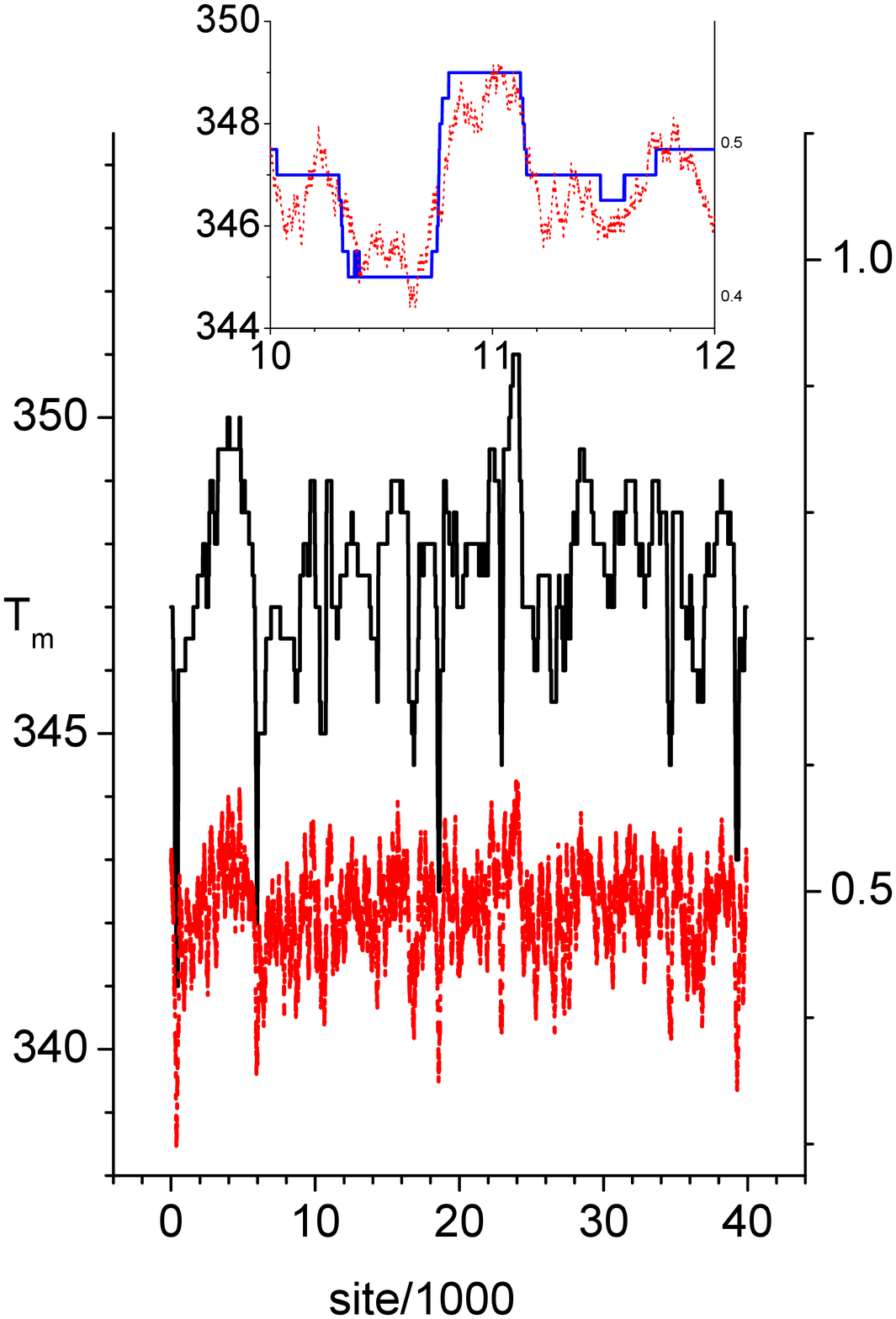,width=6cm }}
%\resizebox{0.5\textwidth}{!}
%\includegraphics{MeltProfN100AT.eps}	
\vspace*{8pt}
\caption {Left panel: The differential melting curve of the T7 phage. The full line shows theoretical results based on the PBD model, the dashed curves experimental results redrawn from \cite{FrankKame1976}. Right panel: the melting map (left y-scale), 200-site moving average of GC content (right y-scale). The inset shows a zoomed region of the melting map. 
}
\label{fig:T7map}
\end{figure}

The procedure described in the previous sections has been used \cite{NTh2010} to compute melting profiles of large genomic sequences. As an  example I show in Fig. \ref{fig:T7map}  the melting profile of the T7 phage, a sequence of 39937 base pairs with a 48.4\% GC content. The experimental data from \cite{FrankKame1976} is also included for comparison. Note that the depths of the Morse wells have been fitted to the data; in fact, this is one of the two sets of data which were used in \cite{NTh2010} to derive the parameters entering Eqs. (\ref{eq:DATpred}).

In order to relate melting profiles to sequence details, it is customary to draw a melting map, in which each site is characterized by its melting temperature, defined via $p_i(T_m(i)=1/2$. The melting map for the T7 phage is shown in the right panel of Fig. \ref{fig:T7map}, along with the moving 200-pt average of local GC-content. 

Two comments are in order here. First the obvious. Genomic heterogeneity destroys the sharp first order transition characteristic of the homogeneous PBD model (and of actual melting profiles of long polynucleotide chains). Remnants of multistep melting can be seen in the inset of Fig. \ref{fig:T7map}, which suggests melting of 200-bp chunks.  This is in general accord with theoretical expectations on the effect of disorder on phase transitions (reduction of effective dimensionality, hence enhancement of the role of fluctuations, rounding of an incipient transition). On the other hand, in an interesting analysis of the effects of heterogeneity within the PBD model context, it has been argued  \cite{CuleHwa} that multistep behavior should be self-averaging, albeit with a large crossover length. I have not detected any systematic signs of self-averaging, even in genomic sequences of much larger lengths than the one presented here.% \cite{}.

The second comment concerns the change in entropy. The left panel of Fig. \ref{fig:T7map} shows the dependence of the entropy derivative on temperature (right y-scale). The curve is almost perfectly superimposed on the melting profile. The total entropy change,  $4.4 k_B$ per base pair, is considerable smaller than the typical experimental values \cite{Blake91} of $12 k_B$ ($24$ cal/mol/bp/K), but represents a considerable improvement from the $1 k_B$ obtained using previous parametrizations \cite{CampaGian} of the PBD model.  

\subsection{Short chains}
\label{seq:short}
\begin{figure}[th]
\centerline{\psfig{file=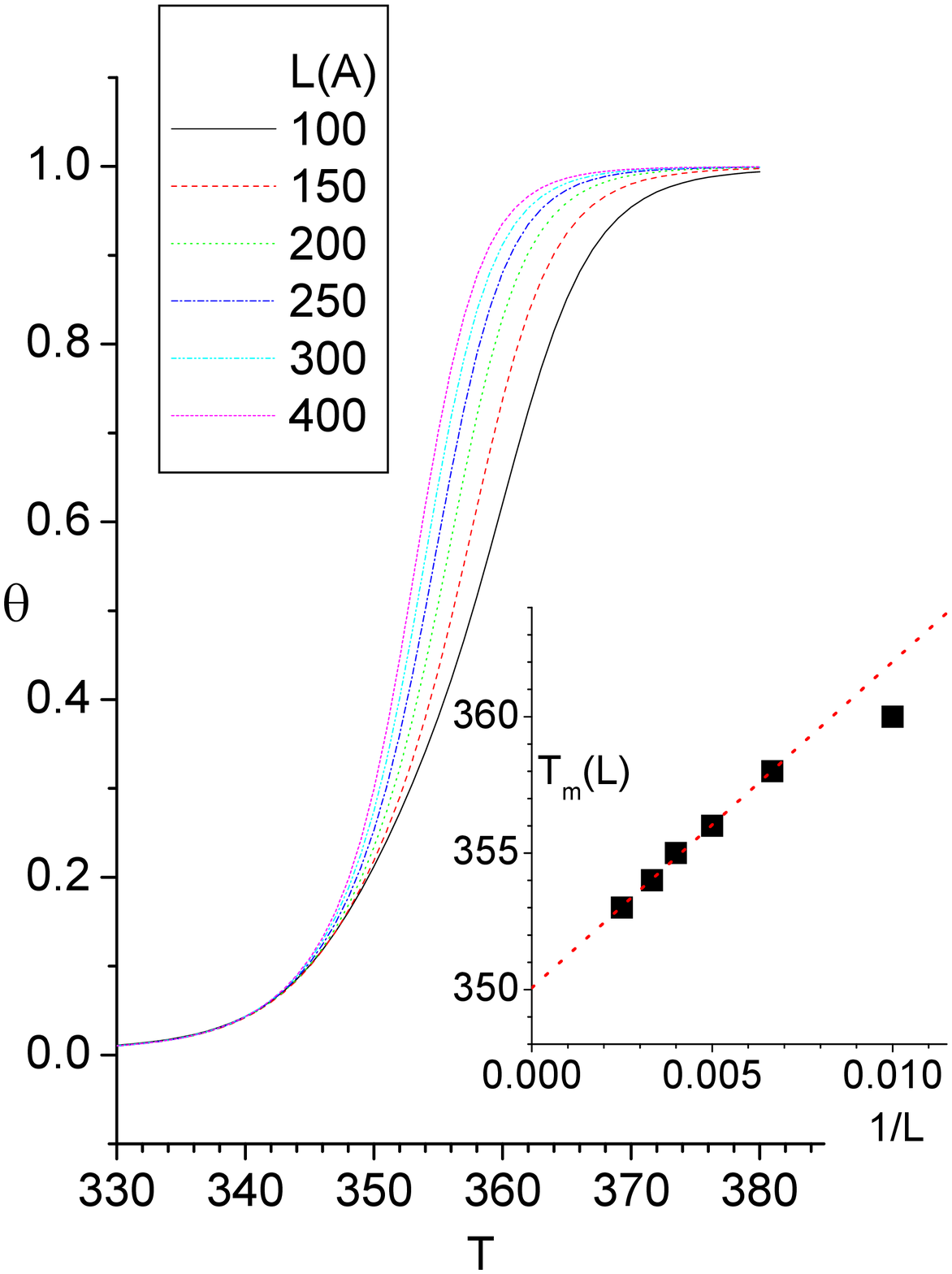,width=5cm} \psfig{file=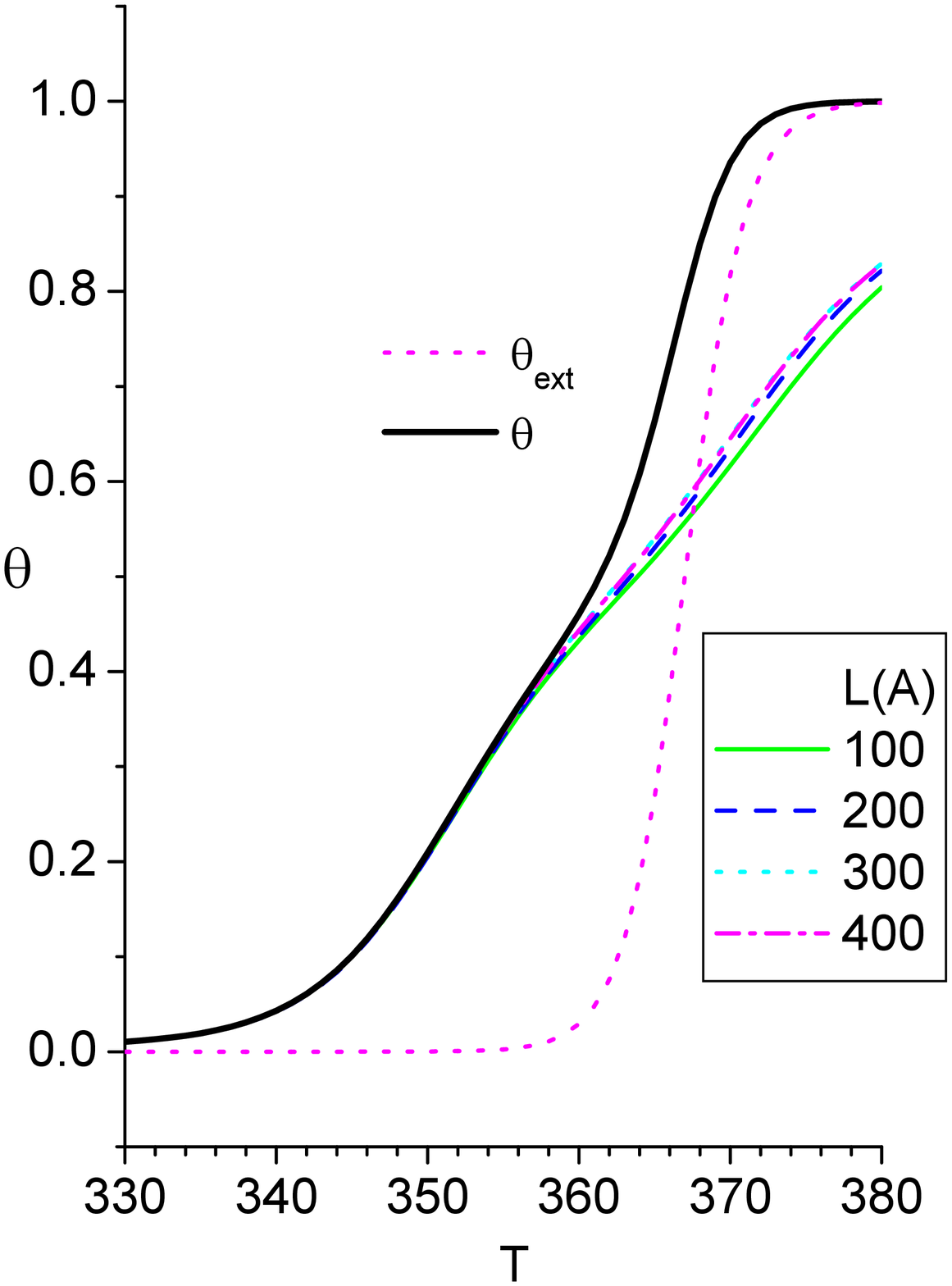,width=5cm } 
\psfig{file=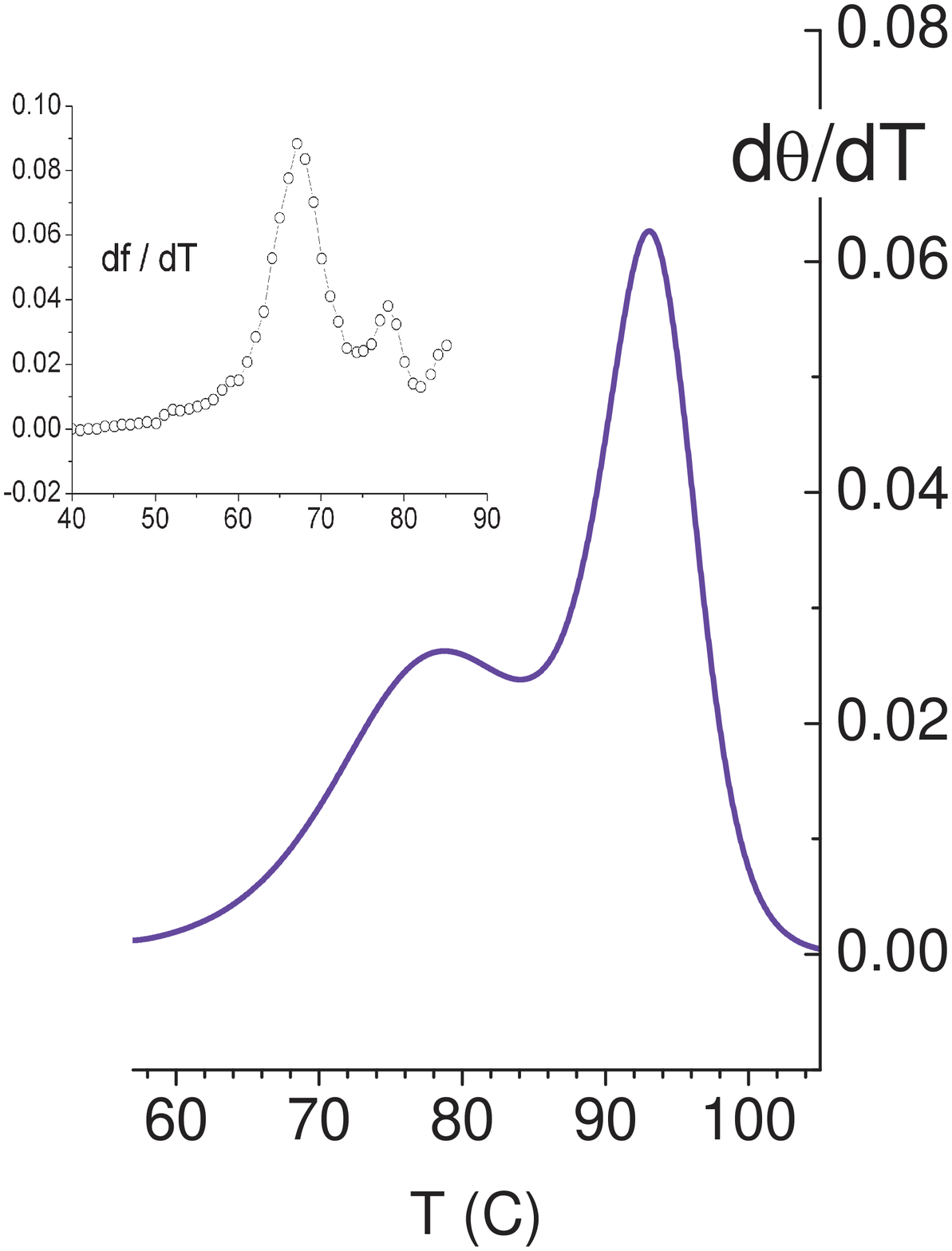,width=5cm }
}	\vspace*{8pt}
\caption {Left panel: The melting curve of the L48AS sequence for a variety of $L$ values. The inset shows the estimated melting temperatures.  
Middle panel: Melting curves in the double-stranded ensemble for a variety of $L$ values. In addition, the dotted curve describes approximately the external melting fraction $\theta_{ext}$. The total fraction of open base pairs is described by the thick solid curve. 
Right panel: the calculated total differential melting curve. Inset: experimental results, reproduced from Fig. 2  of ref. \cite{Montri} with permission. 
}
\label{fig:L48}
\end{figure}

An interesting example of a short chain is the $L48AS$ sequence, which exhibits biphasic melting behavior at $c=50mM$ salt concentration \cite{Montri}.   
Fig. \ref{fig:L48} shows the ($L$-dependent) melting curves in the left panel. 
The curves were obtained using Morse potential depths according to (\ref{eq:DATpred}).
The extracted sequence of melting temperatures (inset, left panel) is regular. Melting behavior is of the single-phase type. This disagrees with experimentally observed behavior, which is not surprising; it should be quite clear that physical separations of the two DNA strands of the order of $300-400$ A should be totally irrelevant to the study a fluctuating 48-bp molecule (which has a total length of 170 A). 

The obvious alternative within the context of the PBD model is the double-stranded ensemble (cf. \ref{seq:dss}). Melting profiles are shown in the middle panel of Fig. \ref{fig:L48}; they are practically $L$-independent. In the same panel I have included an estimate (dotted curve) which mimics the experimentally observed $\theta_{ext}$ \cite{Montri}, in terms of relative temperature position and width. Making use of this estimate it is possible to calculate the total melting profile, which does indeed exhibit biphasic behavior. The differential melting curve is shown in the right panel, along with the corresponding experimental results of Ref.  \cite{Montri}.

In spite of this qualitative agreement, there are still considerable problems with the theoretical description of this exemplary case of biphasic melting in DNA oligomers. First, the predicted peaks in the melting profiles lie more than 10 degrees higher than the experimentally observed ones. Second, the predicted biphasic behavior is of the wrong type, i.e. melting accelerates at higher temperatures (rather than  slowing down, as experimentally observed). This may well signal a fundamental difficulty of describing oligomer melting behavior in terms of the PBD model. 

\section{Statistics of bubbles and clusters}
\subsection{Bubbles}
\lq\lq Denaturation bubbles\rq\rq (here to be referred to in brief as bubbles) are extended regions of space where the displacement exceeds the critical value $y_c$. The probability of having a bubble of length $k$ at site $n$ is  (cf. (\ref{eq:pnmatr}))
\begin{equation}
{\cal Q}(n|k) =  
\frac  {
< {\bf A}^{(1)} |
{\bf B}^{(2)}  
\cdots
{\bf B}^{(n-k)}
{\bf {\cal {\tilde B}}}^{(n-k+1)}
\cdots
{\bf {\cal {\tilde B}}}^{(n)}
{\bf B}^{(n+1)} 
\cdots
 {\bf B}^{(N-1)}
| {\bf A}^{(N)}  >}
{< {\bf A}^{(1)} |
{\bf B}^{(2)}  \cdots {\bf B}^{(N-1)}
| {\bf A}^{(N)}  >
}
 \quad , 
\label{eq:bubble}
\end{equation}
where ${\bf {\cal {\tilde B}}}^{(n)}$ is the complement of  ${\bf {\cal  B}}^{(n)}$, i.e.
\begin{equation}
{\bf {\cal {\tilde B}}}^{(n)} =   {\bf B}^{(n)} -   {{\bf \cal  B}}^{(n)}
\end{equation}
and $n=k+1,\cdots,N-1$, i.e. this is a backward-counted,  \lq\lq internal\rq\rq\-  bubble. In terms of the stored unit vectors and norms of section \ref{sec:heteroTherm}, this can be rewritten in the form
\begin{equation}
{\cal Q}(n|k) =
\frac{1}{ \mu_{n-k+1}^L \cdots \mu_{n}^L  }
\frac  {
< {\bf u}^{(n-k)} |
{\bf {\cal {\tilde B}}}^{(n-k+1)}
\cdots
{\bf {\cal {\tilde B}}}^{(n)}
| {\bf v}^{(N-j)}  >}
{< {\bf u}^{(n)} 
| {\bf v}^{(N-n)}  >
}
 \quad .
\label{eq:normbubble}
\end{equation}
\begin{figure}[th]
\centerline{\psfig{file=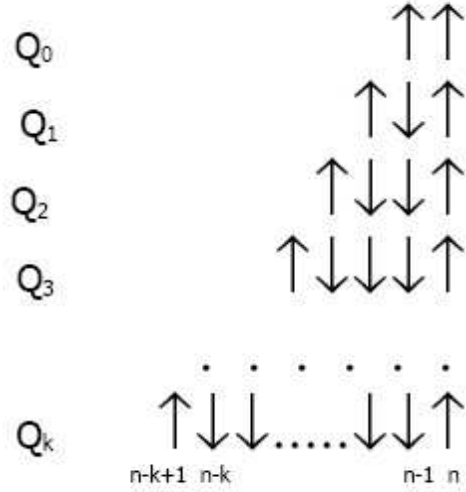,width=7cm}}	
\vspace*{8pt}
\caption { A schematic view of the bubble size probability distribution. Starting at the $n$th site (bound base pair, upward pointing arrow), all possible sequences to the left of $n$ are shown. Either there is no bubble (no downward pointing arrows), or the bubble terminates after $1,2,\cdots k$ sites. Since the list is exhaustive, the sum of all probabilities must equal the probability of the $n$th site being in the bound state.  Similarly, if the first row is excluded, the sum of all remaining probabilities must equal the joint probability of the  $n$th site being in the bound and the $(n-1)$st in the unbound state (flip density).
}
\label{fig:sumschematic}
\end{figure}
Note that a strict definition of a bubble should include the demand that the enclosing sites $n-k$ and $n+1$ should be occupied by bound base pairs. The  probability for this to occur will be
\[{\cal{\hat Q}}(n|k) =
{\cal Q}(n|k) - {\cal Q}(n+1|k+1) - {\cal Q}(n|k+1) + {\cal Q}(n+1|k+2)   
\]
 For very long chains, it is possible to average over all sites ($N'<N$ to exclude boundaries) and obtain the site-averaged  probability 
\begin{equation}
Q_k = \frac{1}{N'} \sum_{n} {\cal{\hat Q}}(n|k)
\label{eq:avprobbub}
\end{equation}
for the occurrence of a (strictly) size $k$ bubble somewhere along the chain. Fig. \ref{fig:sumschematic} provides a 
schematic illustration of the sum rule
\begin{equation}
\sum_{k=0}^{\infty}  Q_k = 1-\theta   \quad.
\label{eq:qsum1}
\end{equation}
In fact, as the figure suggests, the sum rule is valid also for the non-averaged probabilities, i.e.
\begin{equation}
\sum_{k=0}^{\infty}  {\cal{\hat Q}}(n|k) = p_n   \quad \forall n\quad ,
\label{eq:qsum1inh}
\end{equation}
 since the sequences shown exhaust all configurations  in which the $n$th base pair is bound (\lq\lq up\rq\rq).
 
Another quantity of interest is the density of neighboring bound / unbound pairs (spin flips). Going back to Fig.
 \ref{fig:sumschematic} it is straightforward to identify the flip density - which, owing to the topology of the one-dimensional chain, is equal to the average bubble density - with  
\begin{equation}
\sum_{k=1}^{\infty}  Q_k =   1-\theta - Q_0   ,
\label{eq:spinflip}
\end{equation}
where $Q_0$  is the joint probability of two successive sites being in the bound state (the zero-size bubble).

The average size of a bubble can be defined as    
\begin{equation}
\xi_b  = \frac{\sum_{k=1}^{\infty}  k Q_k    }{\sum_{k=1}^{\infty}   Q_k  }  \quad.
\label{eq:avbubsize}
\end{equation}

Noting that the product of average bubble size and density is just the fraction of open base pairs leads to 
a second sum rule, 
\begin{equation}
\sum_{k=1}^{\infty} k Q_k = \theta   \quad.
\label{eq:qsum2}
\end{equation}

The above sum rules (\ref{eq:qsum1}) and (\ref{eq:qsum2})
represent exact properties of the infinite chain. They have been explicitly verified in the 
homogeneous case \cite{NTh2008}. Moreover, since (\ref{eq:qsum1}) can be derived by finite-averaging of
(\ref{eq:qsum1inh}) it will hold exactly for finite, heterogeneous sequences.  (\ref{eq:qsum2}) is 
expected to hold for a long heterogeneous sequence, provided that average bubble number and size 
allow for meaningful statistics.

\begin{figure}[th]
\centerline{\psfig{file=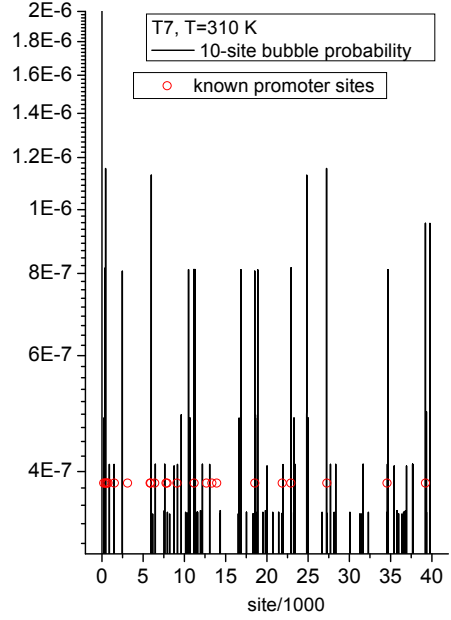,width=7cm}}	
\vspace*{8pt}
\caption {Probability of occurrence of 10-site bubble in the T7-phage. The circles denote published promoter sites \cite{T7prom}.
}
\label{fig:T7bubb}
\end{figure}

As an example of (non-averaged) site bubble distribution in genomic DNA,  
${\cal{\hat Q}}(n|10)$ has been calculated in the case of the T7 phage and  is shown in Fig. \ref{fig:T7bubb}.
Note that the probability of  10-site bubble occurrence can be substantial, i.e. only slightly smaller than the probability of single base-pair opening (estimated to be in the order of 1 ppm by imino proton exchange measurements \cite{Gueron87}).
Clearly, this can only happen in AT-rich regions. The figure includes known \cite{T7prom} promoter sites for (a very qualitative and at the present stage by no means systematic) comparison. 

\subsection{Double-stranded clusters}
\label{sec:clusters}
Equivalently, one may define the probability of a number of successive base pairs being in the bound state. Such a $k$-size double-stranded cluster (in brief: cluster) ending on the $n$th site, and surrounded by open base pairs at the ends, would be associated with a probability
  \[{\cal{\hat P}}(n|k) =
{\cal P}(n|k) - {\cal P}(n+1|k+1) - {\cal P}(n|k+1) + {\cal P}(n+1|k+2)   
\]
where 
 \begin{equation}
{\cal P}(n|k) =
\frac{1}{ \mu_{n-k+1}^L \cdots \mu_{n}^L  }
\frac  {
< {\bf u}^{(n-k)} |
{\bf {\cal  B}}^{(n-k+1)}
\cdots
{\bf {\cal B }}^{(n)}
| {\bf v}^{(N-j)}  >}
{< {\bf u}^{(n)} 
| {\bf v}^{(N-n)}  >
}
 \quad .
\label{eq:normcluster}
\end{equation}
and all the arguments of the previous subsection related to bubbles apply {\em mutatis mutandis} to clusters.
Again, for long genomic sequences, one may perform a site average and obtain the probability of occurrence of a $k$-size cluster anywhere along the chain,
\begin{equation}
P_k = \frac{1}{N'} \sum_{n} {\cal{\hat P}}(n|k)  \quad.
\label{eq:avprobclust}
\end{equation}
The above space-independent probabilities satisfy the sum rules
\begin{eqnarray}
\nonumber
\sum_{k=1}^{\infty} k P_k &=& 1-\theta \\
\sum_{k=0}^{\infty} P_k &=& \theta  \quad.
\label{eq:psums}
\end{eqnarray}
Note that the latter sum includes a quantity $P_0$ which is the (unconditional) probability of two successive sites being in the open state (the zero-size cluster).

The average size of a cluster is 
\begin{equation}
\xi_c  = \frac{1-\theta}{\theta-P_0}  \quad.
\label{eq:avclustsize}
\end{equation}
 The quantity $\theta-P_0$ expresses the probability of finding an open base pair, followed by a closed base pair. Similarly,
the quantity $1-\theta -Q_0$ is equal to the probability of finding a closed base pair, followed by an open base pair. These \lq\lq flip\rq\rq\- probabilities must be equal: 
 \begin{equation}
	1-\theta-Q_0 = \theta - P_0  \quad;
\end{equation}
furthermore, the average length of a combined cluster and the bubble which follows it, is, from (\ref{eq:avbubsize}) and
 (\ref{eq:avclustsize}), 
\begin{equation}
	\xi_c + \xi_b = \frac{1}{\theta -P_0}  \quad,
\end{equation}
i.e. $\theta -P_0=1-\theta -Q_0$ expresses the number density of bubbles and/or clusters.

\begin{figure}[th]
\centerline{\psfig{file=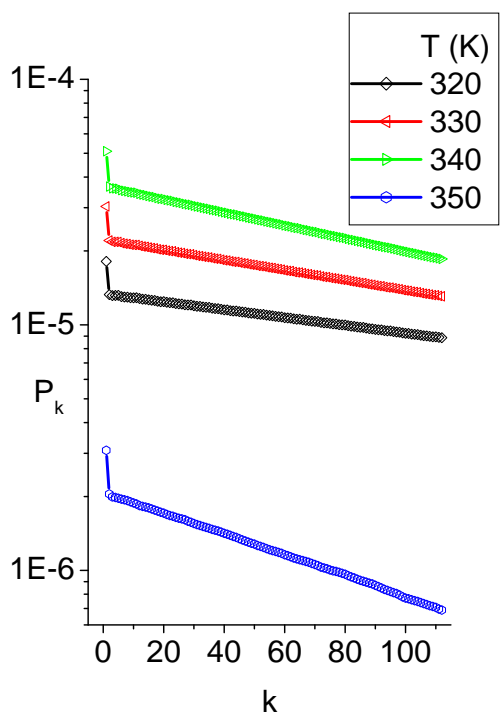,width=7cm}}	
\vspace*{8pt}
\caption {Site-averaged probability of occurrence of a size-$k$ cluster in the T7-phage.
}
\label{fig:T7clust}
\end{figure}

Site-averaged probabilities of cluster occurrence are important in understanding experiments performed with macroscopic samples of genomic DNA, e.g. neutron or X-ray scattering.  Fig \ref{fig:T7clust} shows typical behavior of $P_n$ as a function of $n$ for a variety of temperatures. The sequence used was that of the $T7$ phage. Note the strictly exponential behavior. The slope increases as one approaches melting - and beyond -. Absolute probabilities increase slightly with increasing temperature in the ordered phase since, according to the definition of a strict cluster, it must be surrounded by open sites.
Table  1 %\ref{tab:psums} 
provides a quantitative test of the averaging process in terms of the sum rules (\ref{eq:psums}). Summation beyond the explicitly calculated values ($k=112$) has been extended to infinity using the numerically determined slopes.  
\begin{table}[ht]  %data  8650D
\label{tab:psums}
\tbl{Test of sum rules (\ref{eq:psums}).}
{\begin{tabular}{@{}ccccc@{}} \toprule
T (K) & $\sum k P_k $ & $ 1-\theta $  &  $\sum  P_k $ & $\theta$
% \\ & (Rad/s) & (Rad/s) 
\\ \colrule
320 &    0.9888          & 0.9960 & 0.0040 & 0.0040\\
330 &    0.9889          & 0.9946 & 0.0054 & 0.0054 \\
340 &    0.9847          & 0.9891 & 0.0109  & 0.0109 \\
350 &    0.0216          & 0.0200 & 0.9800 & 0.9800  \\ \botrule
% data files 8650D,8650F  T7_Pn.opj
\end{tabular}}
%\begin{tabnote}
%Table notes
%\end{tabnote}
%\begin{tabfootnote}
%\tabmark{a} Table footnote A\\
%\tabmark{b} Table footnote B
%\end{tabfootnote}
%
\end{table}
\section{Conclusions and Outlook}
Mesoscopic lattice-dynamics modeling of DNA has been described in this paper as a powerful tool for the description of thermal properties. In particular, it has been shown that efficient computational techniques can be used to calculate melting profiles of very long genomic sequences. The resulting - not yet fully optimized - parametrization is in (full or at least partial) accord with results from other independent measurements, e.g. stiffness, Raman spectroscopy, calorimetry, neutron diffraction. On the basis of these findings, it appears that further work is needed in at least three directions. First, a full-scale optimization of model parameters should be undertaken, involving a significantly larger amount of melting data on long chains. Second, an effort should be made to include the computationally intensive, physically important effect of heterogeneous stacking 
\cite{FrankKam2006,BishopNuclAcids2009}
on bubbles in long genomic chains . Third - not necessarily independent of the other two - would be to achieve an improved understanding of the collective (static and dynamic) properties of specially designed oligomers. 
\section*{Acknowledgments}
I would like to thank Professor Giovanni Zocchi (UCLA) and the EDP Sciences Copyright Department (http://epljournal.edpsciences.org) for their kind permission to reproduce Fig. 2 of reference \cite{Montri}.

\end{document}